# Coulomb blockade and Kondo effect in a few-electron silicon/silicon-germanium quantum dot

Levente J. Klein, Donald E. Savage, Mark A. Eriksson
University of Wisconsin-Madison, Madison, WI 53706, USA

Transport measurements at cryogenic temperatures through a few electron top gated quantum dot fabricated in a silicon/silicon-germanium heterostructure are reported. Variations in gate voltage induce a transition from an isolated dot toward a dot strongly coupled to the leads. In addition to Coulomb blockade, when the dot is strongly coupled to the leads, we observe the appearance of a zero bias conductance peak due to the Kondo effect. The Kondo peak splits in a magnetic field, and the splitting scales linearly with the applied field. We also observe a transition from pure Coulomb blockade to peaks with a Fano lineshape.

Quantum dots are ideal systems to probe single electron charging, electron-electron correlations, weak localization, spin effects and cotunneling processes including the Kondo effect.[1-4] Fabrication of highly tunable few electron quantum dots in silicon/silicon-germanium is of interest for spintronics and possible quantum computing applications.[5,6] Much of the interest in silicon from the perspective of these fields stems from the potentially long spin coherence times of electrons in silicon quantum dots. In addition, the presence of multiple conduction band valley minima in silicon is predicted to strengthen coherent spin phenomena such as the Kondo effect in silicon quantum dots.[7,8]

To observe coherent effects such as Kondo in transport measurements on quantum dots requires flexible control of the coupling between the quantum dot and the leads as well as few-electron occupation. Recently, single electron charging in Si/SiGe quantum dots has been observed.[9-11] Here, we report single electron charging in a top-gated quantum dot fabricated in a Si/SiGe heterostructure in the few-electron limit. Tuning the quantum dot so that the tunnel coupling to the leads is strong, we observe both the Kondo and the Fano effects. In the regime in which the dot is well isolated from the leads, the observed Coulomb diamonds are sharp, and we discuss features of the heterostructure that we believe are relevant to this result.

The quantum dot was fabricated in a modulation-doped heterostructure containing an electron gas with a mobility of 77,000 $cm^2/(Vs)$ at 2 K and a carrier density of $4.4 \times 10^{11}$ $cm^{-2}$. The dopant density in the electron supply layer was carefully tuned so that no parallel conduction could be observed in Shubnikov de Haas measurements (data not shown). We have observed a correlation between small slopes for the Shubnikov de Haas magnetoresistance measurements at low to moderate magnetic fields and the ability to fabricate reliable Schottky top-gates capable of defining quantum dots. Top-gates consist of palladium metal deposited on top of a silicon capping layer. Ohmic contacts were formed by evaporation of a Au/Sb film and subsequent annealing at 625 $^oC$ for 30s. An atomic force microscope image of a completed device is shown in the inset to Fig 1a. During the measurements reported in this letter, the voltages on three of the gates were held constant as follows: $V_{g1}$ = -0.450 V, $V_{g2}$ = -0.425 V, and $V_{g3}$ = -0.400 V. To tune the dot potential the voltage $V_{gp}$ was varied. For the gate voltages used, no leakage current was detectable between the gates and the two-dimensional electron gas. Similar quantum dot gate geometries and gate voltage levels have been used in the past to create few-electron quantum dots.[12]

The conductance through the dot measured at the base temperature (20 mK) of a dilution refrigerator is shown in Fig 1a with an applied ac voltage of $V_{sd}$ = 7.5 µV at 9 Hz. The conductance plot displays two well-defined regions as a function of $V_{gp}$: at large negative voltages the dot is well isolated from the leads, and there is zero conductance between the Coulomb blockade peaks. For less negative voltages the dot is strongly coupled to the source and drain leads, leading to the Fano and Kondo effects, as discussed below. At $V_{gp}$ ~ -0.34 V, the dot is sufficiently strongly coupled to the leads that the Coulomb blockade is almost entirely lifted. The measured conductance remains relatively small at these gate voltages because of a large series resistance due to low temperature filters and the device geometry at large distances from the dot.[13]

A two-dimensional plot of the quantum dot conductance is shown in Fig 1b. In the gate voltage regime reported here, small changes in

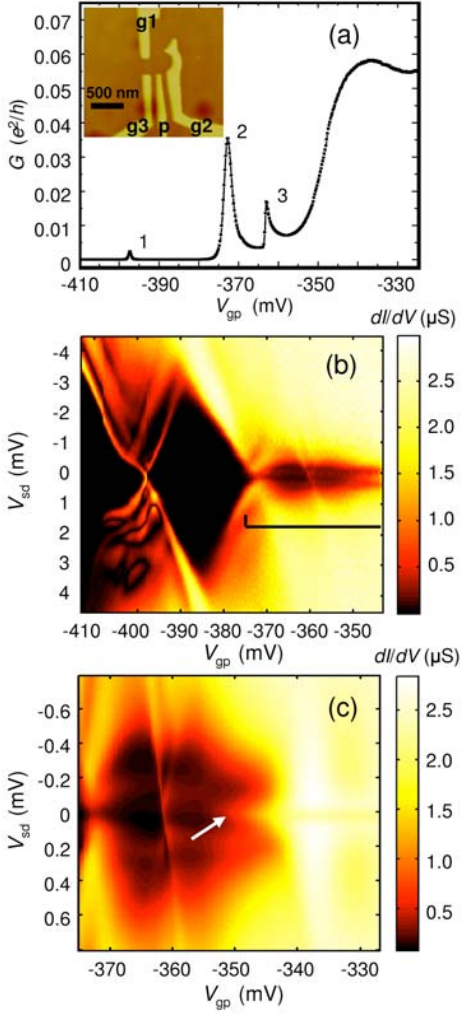

**Fig. 1 (color)** (a) The low-temperature conductance through the quantum dot showing three Coulomb blockade peaks. The third Coulomb peak, at $V_{gp} \sim -363$ mV, has an asymmetric peak shape, the origin of which is discussed in the text. (inset) Atomic force microscope image of the top gated quantum dot with the gates labeled for reference. (b) A two-dimensional plot of the conductance of the quantum dot as a function of the voltage on the central gate $V_{gp}$ and the source-drain bias $V_{sd}$. (c) A high resolution plot of the conductance in the regime of the second and third Coulomb diamonds (as indicated by the square bracket in (b)). Note in particular the sharp resonance lines corresponding to the Coulomb peak at $V_{gp} \sim -363$ mV. The white arrow indicates the position of the Kondo resonance.

$V_{gp}$ cause relatively large changes in the tunnel barrier resistances, leading to two well-defined regions in Fig. 1b. For the first region, with $V_{gp} < -370$ mV, the interior of the Coulomb diamonds have vanishing conductance. For $V_{gp} < -410$ mV, outside the field of view in Fig. 1b, the leftmost diamond continues to be blockaded and no additional Coulomb peaks are observed. The large diamond has sharp, well-defined edges. The slopes of the edges of this diamond allow us to extract a dot capacitance of 42 aF. The gate capacitance for this diamond is 26 aF, and the charging energy is 3.8 meV.

For the second region, with gate voltages in the range -373 mV $< V_{gp} <$ -340 mV, there are two smaller diamonds. In this region there is finite conductance inside the diamonds. Fig 1c shows a high-resolution image of the conductance plot for these two consecutive diamonds. The maximum height of these diamonds in $V_{sd}$ is smaller than for the first large diamond, and their width in $V_{gp}$ is reduced, as is characteristic in few-electron quantum dots. The crossing edges of the Coulomb diamonds in the middle of Fig. 1c are sharp and narrow. Such sharp crossing lines at the edge of Coulomb diamonds have been observed previously and attributed to Coulomb blockade and the Fano effect.[3] In such cases, the interference of a single electron charging transport channel and an additional parallel conduction path leads to the Fano effect. The origin of the second channel cannot be determined from the data, but a recent experiment on dots with similar structure demonstrated how such multiple paths can occur through dots with the gate-design used here.[14] The slopes of the sharp crossing lines in Fig. 1c are almost identical to the slope of the large diamond. The conductance inside of these two diamonds has a finite value, in part due to the appearance of the parallel conduction channel, although single electron charging continues to dominate the transport. This is demonstrated by the Coulomb diamonds themselves and by the similarity of the slopes of the sharp resonance to the slopes of the edges of the largest diamond in Fig. 1b.

The conductance of the Coulomb peak with the Fano effect can be described using the Fano formula:

$$G = G_{incoh} + G_{coh} \frac{|\varepsilon + q|^2}{\varepsilon^2 + 1}, \qquad (1)$$

where the $G_{incoh}$($G_{coh}$) are the incoherent (coherent) contributions to the conductance,

$\varepsilon = \dfrac{2(E - E_0)}{\Gamma}$, $\Gamma$ is the width of the peak, and $E-E_0$ is the shift of the energy from the resonance $E_0$, where E is determined by the gate voltage scaled by the factor $C_g/C$. The correction factor q is the ratio of the transmission amplitude of the resonant and nonresonant channels.[15] The third Coulomb oscillation peak can be fit using Eqn. 1, with the best fit achieved with parameters $q = 0.5347$ and $\Gamma = 130$ μeV, as shown in Fig. 2a.

Upon increasing the temperature in the system, the Coulomb oscillation peaks shown in Fig. 1 display different behaviors. The inset of Fig. 2b shows the first peak, for which increasing the temperature (from base temperature, to 250 mK, to 400 mK) induces a very small increase in the height and width of the peak. For the second peak, small increses in temperature increase both the height and the width of the peak (Fig 2b), and do so more rapidly than for the first peak. For the third peak, in sharp contrast to the first two peaks, the height decreases as the temperature is increased, and the asymmetric peak shape is modified so that the peak becomes increasingly symmetric at elevated temperature (Fig 2b). This transformation from a highly asymmetric peak to a symmetric peak as the temperature is increased is also characteristic of the Fano effect.[3]

As the gate voltage $V_{gp}$ becomes even less negative, the dot-to-lead couplings become strong enough that the Kondo effect appears. The Kondo effect in quantum dots is a second order tunneling process in which a dot electron spin interacts coherently with the Fermi sea in the leads, resulting in an increased density of states (DOS) at the Fermi level.[1] This increased DOS results in a peak in the zero-bias conductance, as shown in Fig. 1c for gate voltages in the range -340 mV < $V_{gp}$ < -350 mV. This enhanced conductance appears as a peak in a plot of the conductance versus $V_{sd}$, as shown in the top curve of Fig. 3a, where we have rescaled the drain-source voltage to take into account the finite contact resistance that acts as a voltage divider.

We investigate the zero bias conductance peak as a function of perpendicular magnetic field $B$, as shown in Fig 3a, from $B = 0$ T to 0.35 T in steps of 0.05 T, from top to bottom. All traces are acquired at gate voltage $V_{gp} = -0.351$ V. At $B = 0$ T, there is a single peak showing enhanced conductance at zero drain source bias. Increasing the magnetic field broadens the peak, and at $B = 0.1$ T the conductance peak has split into a double peak roughly symmetric about $V_{sd} = 0$. Theory of the Kondo effect predicts that the splitting scales linearly with magnetic field and that the splitting is twice the Zeeman splitting.[2] Fig 3b shows a plot of the peak splitting as a function of applied magnetic field. The peak splitting has a linear dependence on magnetic field with a slope of 264 μV/T. The g-factor extracted from this measurement is 2.26, slightly larger than the expected g = 2 in silicon. Orbital effects may play a role in this enhanced value due to the application of the magnetic field perpendicular to the sample.[1]

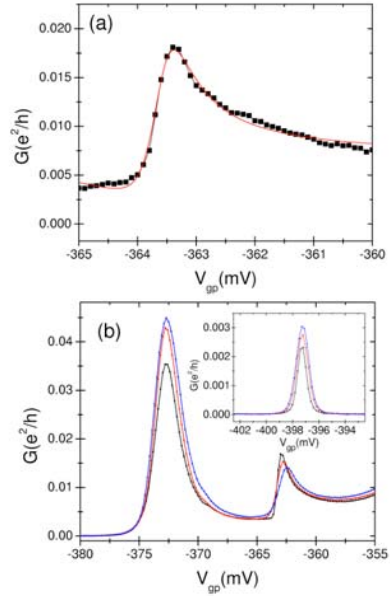

**Fig. 2** (a) A plot of the Coulomb peak (squares) at $V_{gp} \sim -363$ mV. The peak is well fit by a Fano lineshape (solid line). (b) Temperature dependence of the second and third Coulomb peaks for T = 20 mK, 250 mK, and 400 mK. The height and width of the peak at $V_{gp} \sim -373$ mV increase as the temperature is increased. In contrast, the height of the peak with the Fano lineshape decreases with increasing temperature, and the peak becomes increasingly symmetric as the temperature increases. (inset) The temperature dependence of the first Coulomb peak, showing small increases in both height and width as the temperature is increased. The conductance of this peak was measured at the same three temperatures as the main figure.

Increasing temperature destroys the Kondo resonance, resulting in the disappearance of the zero-bias conductance peak at high temperature (Fig. 3c). The conductance peak height decreases as the temperature is increased, and the peak disappears entirely at a temperature around 200 mK. The height of the zero bias conductance peak has a logarithmic dependence on temperature (Fig 3d), a finding that is consistent with previous observations of the Kondo effect in semiconducting quantum dots.[2]

Here we have discussed measurements of Coulomb blockade in a few-electron top-gated Si/SiGe quantum dot. Si/SiGe heterostructure quality has been a key concern for applications such as quantum computing. For example, it has been an open question whether sharp diamonds can be observed in these types of heterostructures. Thus, the observation of sharp diamonds (shown here on the left-hand side of Fig. 1a) demonstrates that heterostructures of sufficient quality can indeed be grown. As discussed above, we believe a key feature in such heterostructures is the dopant density in the electron supply layer. Reliable Schottky gates allowed the dot presented here to be either strongly or weakly coupled to its leads. Strong coupling allowed the observation of both the Kondo and the Fano effects, demonstrating coherent electron transport in a silicon quantum dot. Particularly in the case of the Kondo effect, the observation of a coherent spin phenomenon is a step towards silicon spin qubits for quantum computing applications.

We acknowledge helpful discussions with M. Eto, R. Joynt, M. Friesen, S.N. Coppersmith, and K.A. Slinker, and experimental assistance from L.M. McGuire. This work was supported in part by the NSA and LPS under ARO contract number W911NF-04-1-0389, and by the NSF under Grant No. DMR-0325634.

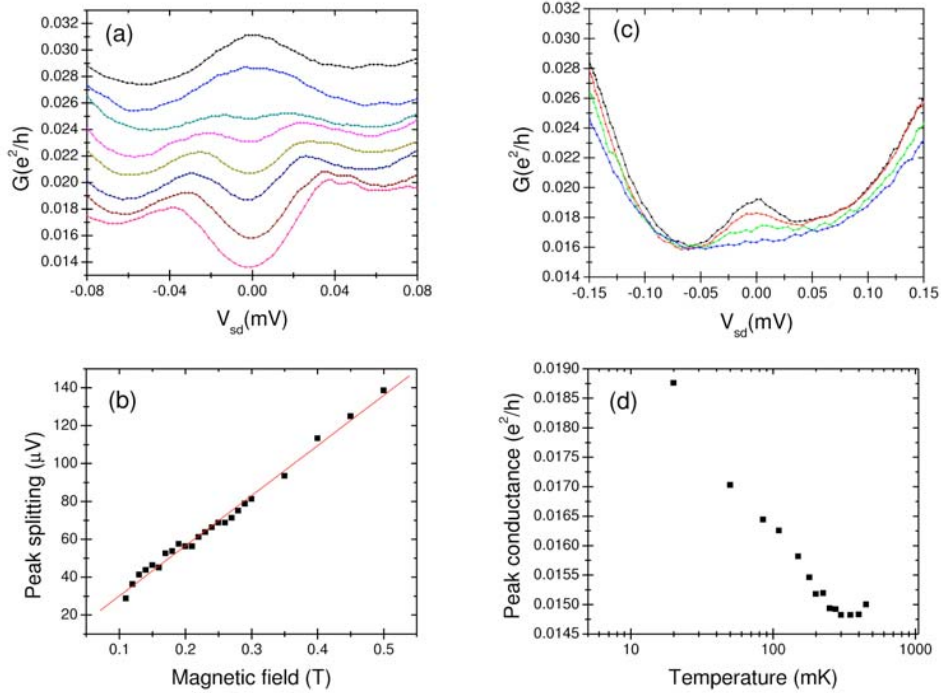

**Fig. 3** (a) The zero bias conductance measured at $V_{gp}$ = -351.1 mV for perpendicular magnetic fields B = 0, 0.05, 0.1, 0.15, 0.2, 0.25, 0.3 and 0.35 T (from top to bottom). The curves are offset for clarity. (b) The extracted peak splitting from (a) as a function of applied magnetic field. The data are well fit by a line with slope 264 µV/T. (c) The temperature dependence of the zero-bias peak height for T = 20 mK, 50 mK, 100 mK, and 200 mK (from top to bottom). (d) Semi-log plot of the peak conductance from (c) as a function of temperature. The height of the zero bias anomaly decreases logarithmically with increasing temperature until the peak is no longer present.